\documentclass{article}

\usepackage[preprint]{neurips_2025}

\usepackage[utf8]{inputenc} 
\usepackage[T1]{fontenc}    
\usepackage{hyperref}       
\usepackage{url}            
\usepackage{booktabs}       
\usepackage{amsfonts}       
\usepackage{nicefrac}       
\usepackage{microtype}      
\usepackage{xcolor}         

\usepackage{soul}
\usepackage{graphicx}
\usepackage{amsmath}
\usepackage{subfigure}
\usepackage{todonotes}
\usepackage{threeparttable}
\usepackage{multirow}
\usepackage[ruled,linesnumbered]{algorithm2e}
\usepackage{array}
\usepackage{bm}
\usepackage{float}
\usepackage{wrapfig}
\usepackage{color}
\usepackage{amssymb}
\usepackage{bbding}
\usepackage{diagbox}
\usepackage{colortbl}
\usepackage{enumitem}

\title{Towards Robust Multimodal Physiological Foundation Models: Handling Arbitrary Missing Modalities}

\author{%
  Wei-Bang Jiang\textsuperscript{1,2}\thanks{Equal contribution}, Xi Fu\textsuperscript{1}$^*$, Yi Ding\textsuperscript{1}\thanks{Corresponding authors}, Cuntai Guan\textsuperscript{1}$^\dagger$ \\
  \textsuperscript{1}Nanyang Technological University\quad\textsuperscript{2}Shanghai Jiao Tong University\\
  \texttt{935963004@sjtu.edu.cn,FUXI0010@e.ntu.edu.sg,\{ding.yi,ctguan\}@ntu.edu.sg} \\
}

\begin{document}

\newcommand{\method}{PhysioOmni\xspace}

\maketitle

\begin{abstract}
    Multimodal physiological signals, such as EEG, ECG, EOG, and EMG, are crucial for healthcare and brain-computer interfaces. While existing methods rely on specialized architectures and dataset-specific fusion strategies, they struggle to learn universal representations that generalize across datasets and handle missing modalities at inference time. To address these issues, we propose PhysioOmni, a foundation model for multimodal physiological signal analysis that models both homogeneous and heterogeneous features to decouple multimodal signals and extract generic representations while maintaining compatibility with arbitrary missing modalities. PhysioOmni trains a decoupled multimodal tokenizer, enabling masked signal pre-training via modality-invariant and modality-specific objectives. To ensure adaptability to diverse and incomplete modality combinations, the pre-trained encoders undergo resilient fine-tuning with prototype alignment on downstream datasets. Extensive experiments on four downstream tasks, emotion recognition, sleep stage classification, motor prediction, and mental workload detection, demonstrate that PhysioOmni achieves state-of-the-art performance while maintaining strong robustness to missing modalities. Our code and model weights are available at \url{https://github.com/935963004/PhysioOmni}.
\end{abstract}
\section{Introduction}
Multimodal physiological signals, such as electroencephalography (EEG), electrocardiography (ECG), electrooculography (EOG), and electromyography (EMG), have garnered increasing interest in brain-computer interfaces (BCI) due to their ability to capture diverse physiological and cognitive states \cite{664275}. EEG reflects neural activity, ECG monitors cardiac rhythms, EOG tracks eye movements, and EMG records muscle activations, collectively offering a comprehensive representation of human physiological responses. These modalities have been widely applied across various domains, including cognitive load assessment \cite{albuquerque2020wauc}, emotion recognition \cite{katsigiannis2017dreamer}, motor imagery \cite{9606552}, and sleep stage classification \cite{khalighi2016isruc}, driving advances in both medical diagnostics and BCI. To enhance performance, numerous algorithms have been proposed to effectively integrate multiple modalities, leveraging their complementary information to improve accuracies across applications \cite{soleymani2011multimodal,chambon2018deep,duan2023dewave,song2024decoding,jiang2025neurolm}.

Over the past two years, several EEG foundation models, including LaBraM, have emerged, demonstrating the feasibility of using masked EEG models for pre-training on large-scale EEG datasets \cite{jiang2024large, wang2024eegpt, wang2025cbramod}. These models have notably improved performance and generalization, laying the groundwork for general unsupervised representation learning. However, a gap persists in the development of general pre-trained foundation models for multimodal physiological signals. Although some studies have explored the integration of multiple physiological signals to boost performance, they are often limited by two key factors: either using multiple modalities during training but relying on a single modality for testing, or being tailored to specific downstream tasks, which restricts their generalizability across diverse datasets. For instance, Fang \textit{et al.} introduced multimodal foundation models designed for sleep stage classification using masked autoencoders \cite{fang2024promoting}, while Brant-X employed contrastive learning for multi-level alignment between EEG and EXG but relied solely on EEG for downstream tasks \cite{zhang2024brant}. Developing a universal multimodal physiological foundation model capable of extracting semantic representations while handling arbitrary missing modalities presents several key challenges:

\textbf{1) Decoupling homogeneous and heterogeneous features}: Multimodal physiological signals encompass both shared and unique patterns. Effectively disentangling modality-invariant (homogeneous) and modality-specific (heterogeneous) features is crucial for robust multimodal learning.

\textbf{2) Unified multimodal representation learning}: Beyond feature decoupling, integrating physiological signals with diverse characteristics to derive effective and generalizable representations remains a fundamental challenge in multimodal representation learning.

\textbf{3) Handling arbitrary missing modalities}: While leveraging all available modalities during training, ensuring robust performance under incomplete modalities at inference time is highly challenging, requiring strategies that maximize adaptability while minimizing performance degradation.

With regard to above challenges, we propose \method, a universal multimodal physiological foundation model pre-trained on diverse multimodal datasets, including EEG, ECG, EOG, and EMG signals. Our approach begins with training a decoupled multimodal tokenizer, where one shared codebook and four private codebooks disentangle multimodal embeddings into modality-invariant and modality-specific codes. These discrete codes serve as the foundation for masked signal modeling, enabling the encoders to learn universal representations across modalities. During fine-tuning, we introduce homogeneous representation mapping, which projects features from different modalities into a common space. To handle arbitrary missing modalities, we incorporate prototype alignment and modality-specific prediction, ensuring robust adaptation across different modality combinations. We comprehensively evaluate \method on four popular BCI tasks: emotion recognition, sleep stage classification, motor prediction, and workload detection. Our approach achieves state-of-the-art (SOTA) performance across both unimodal and multimodal settings, underscoring the effectiveness of decoupled multimodal learning and resilient fine-tuning. Our key contributions are threefold:

\textbf{1) Decoupled multimodal tokenizer}: We design a multimodal tokenizer that disentangles modality-invariant and modality-specific features using a shared codebook for common patterns and private codebooks for unique characteristics, enhancing multimodal fusion.

\textbf{2) Masked signal pre-training}: We extend masked signal modeling to multimodal physiological signals, enabling the model to learn both generic and semantic representations. Leveraging the decoupled tokenizer, we design modality-invariant and modality-specific code prediction, ensuring the model captures both shared and unique characteristics of each signal type.

\textbf{3) Resilient fine-tuning with prototype alignment}: To handle arbitrary missing modalities, we introduce homogeneous representation mapping, prototype alignment, and modality-specific prediction, enabling the model to dynamically adapt to varying modality combinations during inference while maintaining robust performance despite incomplete input signals.

\section{Related Work}
\subsection{Foundation Models for EEG}
EEG foundation models, designed to handle arbitrary configurations while learning robust and versatile representations, have gained significant attention in recent years. BIOT \cite{NEURIPS2023_f6b30f3e} introduces a Biosignal Transformer that tokenizes channels into patches, enabling cross-data learning across heterogeneous biosignal formats (EEG and ECG). LaBraM \cite{jiang2024large} leverages vector-quantized neural spectrum prediction and masked EEG modeling to facilitate cross-dataset learning, significantly enhancing performance across diverse EEG tasks. EEGPT \cite{wang2024eegpt} pre-trains a 10-million-parameter Transformer using mask-based dual self-supervised learning and spatio-temporal representation alignment to improve EEG representation learning. CBraMod \cite{wang2025cbramod} employs a criss-cross Transformer to separately model spatial and temporal dependencies while incorporating an asymmetric conditional positional encoding scheme for greater adaptability.

\subsection{Multimodal Models in BCI}
While many studies incorporate multimodal signals to enhance BCI performance, a universal foundation model that generalizes across modalities and datasets while extracting generic representations remains lacking. VBH-GNN \cite{liu2024vbhgnn} utilizes multimodal physiological signals and variational Bayesian heterogeneous graphs for cross-subject emotion recognition. Brant-X \cite{zhang2024brant} aligns physiological signals by leveraging an EEG foundation model for knowledge transfer but is limited to EEG-based downstream tasks. CIMSleepNet \cite{shen2024robust} addresses arbitrary modality missing in sleep staging through a modal imagination module and contrastive learning, alongside temporal attention for better context representation. However, these models either rely on multimodal inputs during training but evaluate on a single modality or are tailored to specific tasks, limiting their generalizability across datasets.

\begin{figure}[t!]
    \centering
    \includegraphics[width=1\linewidth]{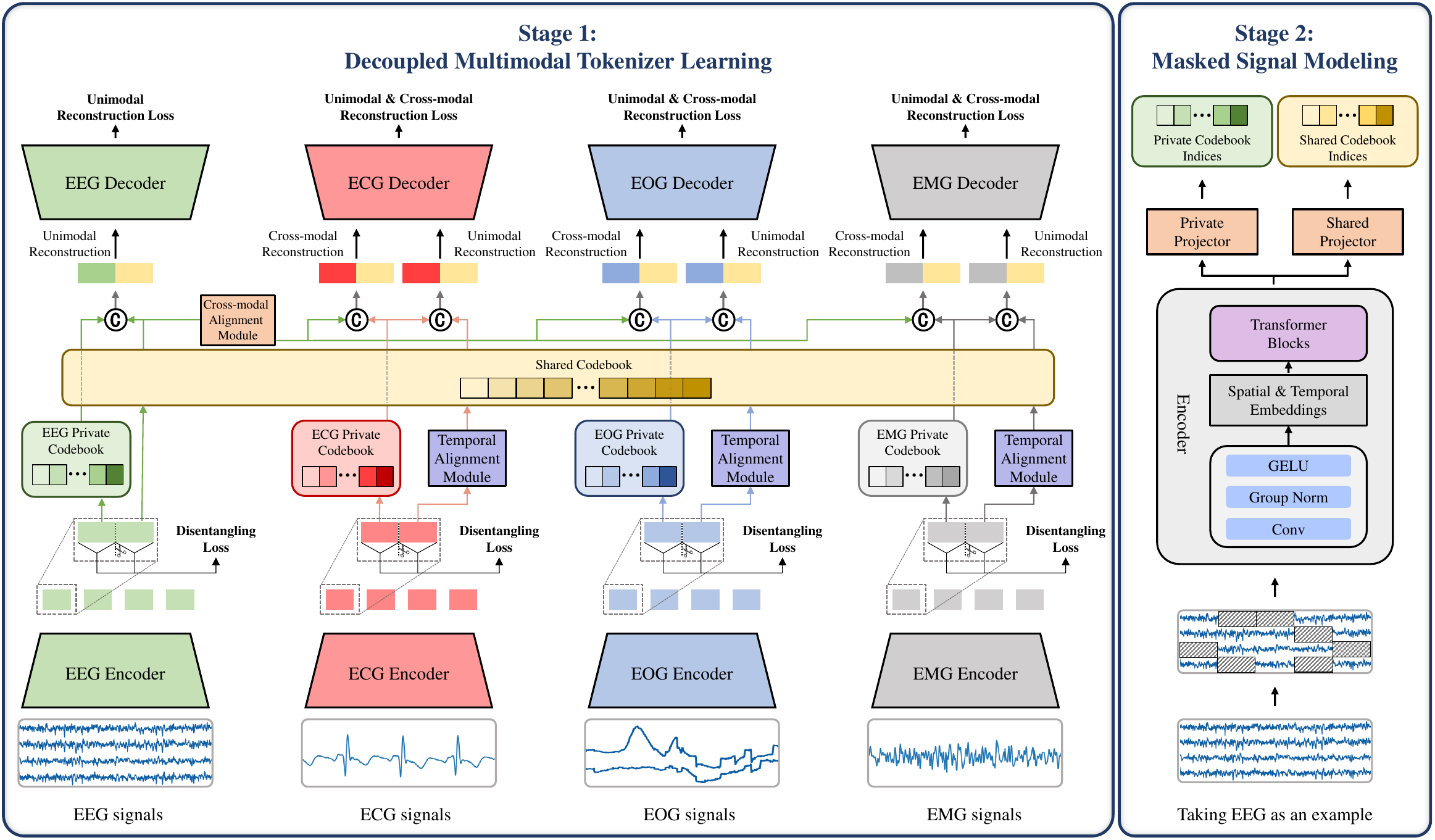}
    \vspace{-15pt}
    \caption{
    Overview of the decoupled multimodal tokenizer learning framework and masked signal modeling. \textbf{Left}: The tokenizer learning framework consists of a shared codebook, private codebooks, encoders, decoders, temporal alignment modules, and cross-modal alignment modules. \textbf{Right}: The masked signal modeling process includes two projectors for private and shared code prediction.
    }
    \label{pretrain}
\end{figure}

\section{Method}
In essence, \method consists of three training stages: 1) Joint learning of a shared codebook and four modality-specific private codebooks through decoupled multimodal tokenizer training; 2) Masked signal modeling to learn generic representations by predicting private and shared codes from masked inputs; 3) Resilient fine-tuning of pre-trained encoders with self-alignment to handle arbitrary missing modalities in downstream tasks. Figure~\ref{pretrain} shows the first two stages.

Given paired multimodal samples $X=\{(x^e_i,x^c_i,x^o_i,x^m_i)\}^N_{i=1}$, where $x^j_i\in\mathbb{R}^{C_j\times T_j}$, with $C_j$ and $T_j$ representing the number of channels and time points, we segment the signals into patches ${x}_i^{j}\in\mathbb{R}^{N_j\times P_j}$ before passing them into the networks, where $N_j=\frac{C_jT_j}{P_j}$ represents the number of patches and $P_j$ is the patch size for modality $j$. We denote the sampling rate for modality $j$ as $S_j$.

\textbf{Model Architecture.} We adopt the encoder architecture of LaBraM \cite{jiang2024large} for all modalities due to its effectiveness and simplicity. A lightweight temporal encoder, consisting of several 1-D convolutional layers followed by GroupNorm \cite{Wu_2018_ECCV} and GELU \cite{hendrycks2016gaussian} activation, is used to extract temporal features within each patch. To incorporate temporal and channel-specific information for each modality, learnable temporal and and spatial embeddings are added to the extracted features before feeding them into the Transformer blocks. Unlike LaBraM, we incorporate RMSNorm \cite{zhang2019root} and SwiGLU \cite{shazeer2020glu} into the Transformer blocks, as done in LLaMA \cite{touvron2023llama}. The same backbone is used for both multimodal tokenizer training and masked signal modeling.

\subsection{Decoupled Multimodal Tokenizer Training}
At this stage, we aim to extract decoupled, compact, and semantically meaningful representations from multimodal physiological signals to facilitate subsequent masked signal modeling in pre-training. To achieve this, we first define four modality-specific encoders $\mathcal{E}^e,\mathcal{E}^c,\mathcal{E}^o,\mathcal{E}^m$ for EEG, ECG, EOG, and EMG, respectively, to extract both private and shared latent embeddings from the input signals:
\begin{equation}
    z_i^{ep},z_i^{es}=\mathcal{E}^e(x^e_i)\quad z_i^{cp},z_i^{cs}=\mathcal{E}^c(x^c_i)\quad z_i^{op},z_i^{os}=\mathcal{E}^o(x^o_i)\quad z_i^{mp},z_i^{ms}=\mathcal{E}^m(x^m_i),
\end{equation}
where $z_i^{jp}$ and $z_i^{js}$ represent the private and shared embeddings for modality $j$, obtained by splitting the encoder outputs. Unlike prior studies \cite{10.1145/3394171.3413678,Li_2023_CVPR,xia2023achieving} that rely on separate encoders for decoupled embeddings, we employ a single encoder per modality to reduce computational overhead.

\textbf{Codebook Optimization.} We design five learnable codebooks: one shared codebook ($\mathcal{V}^s\in\mathbb{R}^{K\times D}$) and four private codebooks ($\mathcal{V}^j\in\mathbb{R}^{K\times D}$ for modality $j$), where $K$ is the codebook size and $D$ is the code dimension. These codebooks extract both modality-invariant and modality-specific features from multiple modalities. Notably, the shared codebook operates at the largest temporal scale (EEG). The shared and private embeddings from the encoders retrieve the closest codebook entry by looking up their respective codebooks and replacing themselves with the nearest code:
\begin{equation}
    \hat{z}_i^{jp}=\mathcal{V}^{j}[\mathop{\arg\min}\limits_k\Vert \ell_2(z_i^{jp})-\ell_2(v_k^j)\Vert_2]\quad\hat{z}_i^{js}=\mathcal{V}^s[\mathop{\arg\min}\limits_k\Vert \ell_2(\text{TA}(z_i^{js}))-\ell_2(v_k^s)\Vert_2],
\end{equation}
where $v_i\in\mathcal{V}$ is the code and $k\in[1,K]$. To enhance stability, we apply $\ell_2$ normalization to the embeddings, which is equivalent to selecting codes based on cosine similarity. Since the temporal scale (patch size) differs across modalities, the Temporal Alignment module (TA) is crucial for aligning smaller-scale modalities with EEG. It is implemented by a cross-attention layer, where a query aggregates $\frac{P_j}{P_e}$ patches into a single embedding:
\begin{equation}
    \text{TA}(z_i^{js})=\text{CrossAttention}(q,W^K_jz_i^{js},W^V_jz_i^{js}).
\end{equation}
where $q\in\mathbb{R}^{1\times D}$ is a learnable query, and $W^K_j$ and $W^V_j$ are the key and value projection weights.

Given the unique characteristics of each modality, EEG and EMG signals are crucial in the frequency domain \cite{dressler2004awareness,komi1979emg}, while EOG reflects eye movements, and ECG exhibits periodic patterns. We propose reconstructing the Fourier amplitude for EEG and EMG signals, while preserving the original signals for EOG and ECG. The codes from the private and shared codebooks are first $\ell_2$ normalized, concatenated, and then fed into the decoders $\mathcal{D}^e,\mathcal{D}^c,\mathcal{D}^o,\mathcal{D}^m$ to reconstruct the target signals:
\begin{equation}
    o_i^e=\mathcal{D}^e(\hat{z}_i^{ep}\|\hat{z}_i^{es})\quad o_i^c=\mathcal{D}^c(\hat{z}_i^{cp}\|\hat{z}_i^{cs})\quad o_i^o=\mathcal{D}^o(\hat{z}_i^{op}\|\hat{z}_i^{os})\quad o_i^m=\mathcal{D}^m(\hat{z}_i^{mp}\|z_i^{ms}),
\end{equation}
where $o_i$ denotes the reconstructed signals, and $\|$ represents concatenation operator. The training loss for codebook optimization is then formulated as:
\begin{equation}
    \mathcal{L}_{CB}=\sum_i(\sum_{j}\underbrace{\Vert o_i^j-x_i^j\Vert_2^2}_{\text{reconstruction loss}}+\sum_{l}(\underbrace{\Vert\bm{sg}(\ell_2(z_i^l))-\ell_2(\hat{z}_i^l)\Vert_2^2}_{\text{VQ loss}}+\underbrace{\Vert\ell_2(z_i^l)-\bm{sg}(\ell_2(\hat{z}_i^l))\Vert_2^2}_{\text{commitment loss}})),
\end{equation}
where $j\in\{e,c,o,m\}$ and $l\in\{e,c,o,m,s\}$. The stop-gradient operation is denoted as $\bm{sg}$. In this framework, the decoders are optimized via the reconstruction loss, the codebooks are updated using the VQ loss, and the encoders are refined through the commitment loss. We apply z-score normalization to the reconstruction target in each sample to enhance the stability of convergence. Additionally, we employ an exponential moving average strategy to ensure stable codebook updates\cite{van2017neural}.

\textbf{Cross-modal Reconstruction.} EEG signals encapsulate rich cognitive and physiological information, as they directly reflect neural activities \cite{da1991neural}. To encourage the shared codebook to capture common patterns across modalities, we designate EEG as an anchor and leverage its shared embeddings to reconstruct the other modalities with their private codes:
\begin{equation}
    \overline{o}_i^c=\mathcal{D}^c(\hat{z}_i^{cp}\|\text{CMA}(\hat{z}_i^{es}))\quad \overline{o}_i^o=\mathcal{D}^o(\hat{z}_i^{op}\|\text{CMA}(\hat{z}_i^{es}))\quad \overline{o}_i^m=\mathcal{D}^m(\hat{z}_i^{mp}\|\text{CMA}(\hat{z}_i^{es})),
\end{equation}
where CMA denotes the Cross-modal Alignment module. Similar to TA, 
CMA employs cross-attention to handle varying temporal scales across modalities. It extends the shared EEG codes by a factor of $\frac{P_eS_j}{P_jS_e}$ to match the patch count of each modality $j\in\{c,o,m\}$: $\text{CrossAttention}(Q,W^K_jz_i^{js},W^V_jz_i^{js})$, where $Q\in\mathbb{R}^{\frac{P_eS_j}{P_jS_e}\times D}$ is the learnable query matrix. The cross-modal reconstruction loss is then computed as:
\begin{equation}
    \mathcal{L}_{CR}=\sum_i \sum_{j}\Vert \overline{o}_i^j-x_i^j\Vert_2^2.
\end{equation}

\textbf{Disentangling Loss.} Beyond ensuring that the shared codebook captures modality-invariant information, it is equally crucial to distinguish shared and private embeddings, ensuring they encode distinct aspects of the input. To mitigate redundancy between embeddings, we introduce a disentangling loss that enforces soft orthogonality:
\begin{equation}
    \mathcal{L}_{D}=\sum_i\sum_j\text{sim}(z_i^{jp},z_i^{js}),
\end{equation}
where cosine similarity is used to encourage orthogonality.

Ultimately, the total loss for decoupled multimodal tokenizer training integrates these constraints:
\begin{equation}
    \mathcal{L}_{T}=\mathcal{L}_{CB}+\alpha_1\mathcal{L}_{CR}+\alpha_2\mathcal{L}_{D},
\end{equation}
where $\alpha_1$ and $\alpha_2$ are weighting factors balancing the loss terms.

\begin{figure}[t!]
    \centering
    \includegraphics[width=1\linewidth]{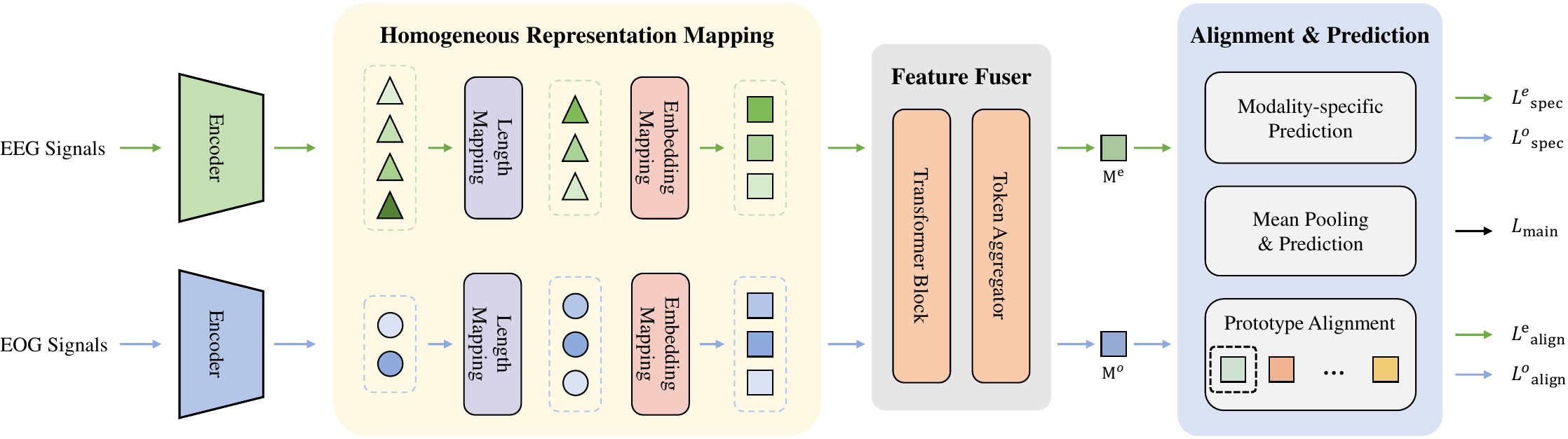}
    \vspace{-15pt}
    \caption{Schematic of the resilient fine-tuning process, illustrated with EEG and EOG as examples. Outputs from all modality encoders undergo Homogeneous Representation Mapping, followed by a Feature Fuser, which consists of a Transformer Block and a Token Aggregator for multimodal feature integration. Prototype alignment ensures robustness to missing modalities in downstream tasks, while modality-specific prediction preserves the performance of individual modalities.}
    \label{finetune}
\end{figure}

\subsection{Masked Signal Pre-training}
Masked signal modeling, an effective representation learning paradigm across various domains \cite{devlin2019bert, he2022masked, jiang2024large}, is extended here to learn generic and semantic representations for multimodal signals. For each modality $j$, we randomly generate a mask $\mathcal{M}=\{0,1\}^{N_j}$ with a mask ratio $r_j$. All masked patches are replaced by a learnable mask token, producing a corrupted sample $\widetilde{x}_i^j$. The encoder is then trained to predict the corresponding codebook indices for each masked patch. Since the embeddings are decoupled into shared and private codebooks, we employ two projection heads to recover the masked patches. Due to varying temporal scales, shared codes may span multiple patches. To ensure alignment, we duplicate the shared codes $\frac{P_eS_j}{P_jS_e}$ times. The training objective is formulated as:
\begin{equation}
    \mathcal{L}_M=-\sum_i\sum_{m_k=1,m_k\in\mathcal{M}}\log p(\hat{z}_i^{jp},\hat{z}_i^{js}|\widetilde{x}_i^j).
\end{equation}

\subsection{Resilient Fine-tuning with Prototype Alignment}
In this stage, we aim to maintain the representational capacity of individual modalities while addressing the challenge of missing modalities. Figure~\ref{finetune} provides an overview of this process. 

\textbf{Homogeneous Representation Mapping.} Each modality-specific input $x^j_i$ is first encoded using its respective pre-trained encoder $\mathcal{E}^j$, producing feature representations $z_i^j\in\mathbb{R}^{n^j\times d^j}$, where $n^j$ and $d^j$ denote the sequence length and dimension, respectively. Since different modalities yield features of varying lengths and dimensions, we introduce Homogeneous Representation Mapping to project them into a unified feature space $\mathbb{R}^{n\times d}$ \cite{zhang2023learning}. 

Following encoding, a modality-specific Length Mapping module $\ell_m$ standardizes feature lengths by transforming the extracted representations into a common-length format $f^j_i\in\mathbb{R}^{n\times d^j}$:
\begin{equation}
    f^j_i=\ell_m(z^j_i)=\mathcal{R}(z^j_i,\widetilde{z}^j_i),\quad\widetilde{z}^j_i=\text{softmax}(\mathcal{H}^j(z^j_i))\in\mathbb{R}^{n^j\times n},\quad\mathcal{R}(a,b)=b^Ta.
\end{equation}
Here, the encoded representations $z^j_i$ are first transformed by a modality-specific head $\mathcal{H}^j$, which consists of linear and normalization layers, to produce $\widetilde{z}^j_i$, a probability distribution over spatial locations obtained via a softmax operation. To facilitate cross-modality alignment, a Reorganization Function $\mathcal{R}$ restructures spatial distributions based on the encoded feature representations. Finally, a modality-specific Embedding Mapping module $e_m$ projects the reorganized embeddings into a common feature space using a linear transformation: $\hat{f}^j_i=e_m(f^j_i)=\text{Linear}(f^j_i)\in\mathbb{R}^{n\times d}$.

\textbf{Fusion \& Alignment.} Given the encoded feature $\hat{f}^j_i$, the Feature Fuser module integrates information from multiple tokens using a shared Transformer block, followed by a Token Aggregator, which applies a 1D convolution layer per modality to aggregate sequential information:
\begin{equation}
h^j_i=\text{Aggregator}(\text{Transformer}(
\hat{f}^j_i))\in\mathbb{R}^{d}.
\end{equation}

To enhance robustness against missing modalities while ensuring effective modality fusion, we introduce three losses in the Alignment \& Prediction module. First, to align unimodal features within a common space, we propose Prototype Alignment, where a set of learnable prototypes $\mathcal{U}=\{u_1,...,u_{|\mathcal{U}|}\}\in\mathbb{R}^{|\mathcal{U}|\times d}$ is shared across all modalities. Specifically, for an $n$-class classification task, each sample is encouraged to be close to the unique prototype corresponding to its class, setting $|\mathcal{U}|=n$. For regression tasks, $|\mathcal{U}|$ is a hyperparameter, and each sample is encouraged to align with the nearest prototype. Therefore, the alignment loss is defined as:
\begin{equation}
    \mathcal{L}_{\text{align}}=\sum_i\sum_j\left\|h^j_i-u_{h_i}\right\|_2^2,
\end{equation}
where $u_{h_i}=\mathcal{U}[\text{label}(i)]$ (classification), $u_{h_i}=\mathcal{U}[\mathop{\arg\min}\limits_k\Vert\ell_2(h_i^j)-\ell_2(u_i)\Vert_2]$ (regression). This loss ensures that the feature representations $h^j_i$ align with their respective prototypes.

The main prediction loss $\mathcal{L}_{\text{main}}$ is computed by averaging the features $h^j_i$ across all modalities, reinforcing the effectiveness of multimodal fusion, which is also used as the final prediction at test stage. Additionnaly, a modality-specific loss $\mathcal{L}^j_{\text{spec}}$ is introduced to preserve the independent representational capacity of each modality. For $\mathcal{L}_{\text{main}}$ and $\mathcal{L}^j_{\text{spec}}$, we employ cross-entropy loss with label smoothing for multi-class classification, binary cross-entropy (BCE) loss for binary classification, and mean squared error (MSE) loss for regression tasks. The total fine-tuning loss is a weighted sum of all components:
\begin{equation}
    \mathcal{L}_{F}=\gamma_m\mathcal{L}_\text{main}+\sum_j\gamma_j\mathcal{L}^j_\text{spec}+\gamma_a\mathcal{L}_\text{align},
\end{equation}
where $\gamma$ values control the trade-off between different loss terms.

\section{Experiments}
\subsection{Downstream Datasets}
\label{downstreamdatasets}
We consider 4 popular downstream tasks in BCI for evaluating \method, where EEGMAT is presented in Appendix~\ref{EEGMAT}. The detailed information is presented in Table~\ref{tab:downstreamdataset}:
\begin{itemize}
    \item \textbf{SEED-VII} \cite{10731546} (emotion recognition): SEED-VII features seven emotions (happiness, sadness, neutral, fear, disgust, surprise, and anger) of 20 subjects who underwent 4 sessions on different days. Each session contains 20 trials of video clips lasting 2-5 minutes. For data partitioning, all subjects' data are combined, with the first 10 trials designated as the training set, the middle 5 trials as the validation set, and the final 5 trials as the test set.
    \item \textbf{HMC} \cite{10.1371/journal.pone.0256111} (sleep stage classification): HMC was developed for automatic sleep scoring, focusing on five sleep stages: wake, NREM-1, NREM-2, NREM-3, and REM. The dataset consists of randomly selected patient recordings from a diverse population undergoing polysomnographic (PSG) examinations for various sleep disorders. It includes full-night polysomnographic recordings from 151 subjects, with the first 100 used for training, the next 25 for validation, and the remaining 26 for testing.
    \item \textbf{FBM} \cite{brantley2018full} (motor prediction): This dataset includes full-body motion capture (66 targets) from approximately 10 walking trials conducted by 10 able-bodied individuals during gait tasks on level ground, ramps, and stairs, with the advantage of unconstrained movement using a motion capture system with wireless IMUs. We use a data stride of 50 ms. For model evaluation, data from all subjects are combined, with the last trial used for testing, the second-to-last trial for validation, and the remaining trials for training.
\end{itemize}

\begin{table}[H]
\centering
\caption{Information of datasets used for downstream evaluation.}
\resizebox{\textwidth}{!}{
\begin{tabular}{@{}lccccc@{}}
\toprule
{\textbf{Dataset}} & {\textbf{Modality (\#Channel)}} & {\textbf{Sampling Rate}} & {\textbf{Duration}} & {\textbf{\#Sample}} & {\textbf{Task}}\\
\midrule
SEED-VII & EEG (62), EOG, ECG & 1000 Hz & 1 second & 281,679 & 7-class classification \\
HMC & EEG (4), EOG, EMG & 256 Hz & 30 seconds & 137,243 & 5-class classification \\
FBM & EEG (60), EOG, EMG (12) & 1000 Hz & 2 seconds & 166,028 & Regression \\
EEGMAT & EEG (19), ECG & 500 Hz & 4 seconds & 2,088 & Binary classification \\
\bottomrule
\end{tabular}
}
\label{tab:downstreamdataset}
\end{table}

\subsection{Experimental Setup}
\label{setup}
\textbf{Preprocessing \& Basic Settings.} All modalities undergo a consistent preprocessing pipeline comprising a bandpass filter, a notch filter, and resampling. The specific parameters are tailored to each modality's characteristics: EEG (bandpass: 0.1–75 Hz, notch: 50 or 60 Hz, resampling: 200 Hz), EOG (bandpass: 0.1–75 Hz, notch: 50 or 60 Hz, resampling: 200 Hz), ECG (bandpass: 0.5–60 Hz, notch: 50 or 60 Hz, resampling: 500 Hz), and EMG (bandpass: 5–200 Hz, notch: 50, 100, 150 Hz or 60, 120, 180 Hz, resampling: 500 Hz). Signals are scaled to $\mu V$ units, and all values are divided by 100 for normalization. The patch size is set to 200 for EEG (1 second) and 100 for ECG (0.2 second), EOG (0.5 second), and EMG (0.2 second). This preprocessing largely follows existing methods \cite{jiang2024large,jin2025reading}, which are widely adopted. Additional details on hyperparameter settings are provided in Appendix~\ref{hyperparameters}.

\textbf{Training Data \& Environment Settings.} To enhance the model's generalization ability, we extensively gather diverse multimodal physiological data from multiple datasets, mostly containing three or more modalities, as detailed in Appendix~\ref{data}. Our experiments are conducted on four NVIDIA A100-80G GPUs with Python 3.12.8 and PyTorch 2.5.1 + CUDA 12.4. The best models are selected based on their performance on the validation set and subsequently evaluated on the test set with full modalities. To ensure reliability and comparability, we report the average and standard deviation across three random seeds.

\begin{table}[H]
\centering
\caption{The results of different methods on SEED-VII.}
\resizebox{\textwidth}{!}{
\begin{tabular}{@{}lccccc@{}}
\toprule
\textbf{Method} & \textbf{Training Modality} & \textbf{Test Modality} & \textbf{Balanced Accuracy} & \textbf{Cohen's Kappa} & \textbf{Weighted F1} \\
\midrule
EEG-Conformer \cite{9991178} & EEG & EEG & \cellcolor{blue!25}0.1775$\pm$0.0014 & \cellcolor{blue!25}0.0428$\pm$0.0015 & \cellcolor{blue!25}0.1482$\pm$0.0030 \\
\multirow{3}*{BIOT \cite{NEURIPS2023_f6b30f3e}} & EEG & EEG & \cellcolor{blue!25}0.3151$\pm$0.0047 & \cellcolor{blue!25}0.2044$\pm$0.0054 & \cellcolor{blue!25}0.3210$\pm$0.0055 \\
& EOG & EOG & \cellcolor{yellow!25}\underline{0.1888}$\pm$0.0015 & \cellcolor{yellow!25}\underline{0.0529}$\pm$0.0015 & \cellcolor{yellow!25}\underline{0.1656}$\pm$0.0042 \\
& ECG & ECG & \cellcolor{red!25}\underline{0.2186}$\pm$0.0015 & \cellcolor{red!25}\underline{0.0925}$\pm$0.0021 & \cellcolor{red!25}\underline{0.2212}$\pm$0.0033 \\
LaBraM-Base \cite{jiang2024large} & EEG & EEG & \cellcolor{blue!25}\underline{0.3456}$\pm$0.0028 & \cellcolor{blue!25}\textbf{0.2391}$\pm$0.0025 & \cellcolor{blue!25}\textbf{0.3511}$\pm$0.0027 \\
CBraMod \cite{wang2025cbramod} & EEG & EEG & \cellcolor{blue!25}0.3237$\pm$0.0026 & \cellcolor{blue!25}0.2126$\pm$0.0028 & \cellcolor{blue!25}0.3270$\pm$0.0026 \\
\multirow{2}*{Fu \textit{et al.} \cite{10340491}} & EEG+EOG & EOG & \cellcolor{yellow!25}0.1449$\pm$0.0017 & \cellcolor{yellow!25}0.0025$\pm$0.0021 & \cellcolor{yellow!25}0.0428$\pm$0.0069 \\
& EEG+ECG & ECG & \cellcolor{red!25}0.1703$\pm$0.0054 & \cellcolor{red!25}0.0337$\pm$0.0066 & \cellcolor{red!25}0.1148$\pm$0.0129 \\
FeatFusion & EEG+EOG+ECG & EEG+EOG+ECG & \cellcolor{green!25}0.3475$\pm$0.0047 & \cellcolor{green!25}0.2301$\pm$0.0045 & \cellcolor{green!25}0.3437$\pm$0.0042 \\
\midrule
\multirow{7}*{\method} & \multirow{7}*{EEG+EOG+ECG} & EEG & \cellcolor{blue!25}\textbf{0.3479}$\pm$0.0054 & \cellcolor{blue!25}\underline{0.2316}$\pm$0.0045 & \cellcolor{blue!25}\underline{0.3442}$\pm$0.0025 \\
& & EOG & \cellcolor{yellow!25}\textbf{0.2079}$\pm$0.0052 & \cellcolor{yellow!25}\textbf{0.0737}$\pm$0.0051 & \cellcolor{yellow!25}\textbf{0.2104}$\pm$0.0084 \\
& & ECG & \cellcolor{red!25}\textbf{0.2302}$\pm$0.0049 & \cellcolor{red!25}\textbf{0.1062}$\pm$0.0065 & \cellcolor{red!25}\textbf{0.2425}$\pm$0.0058 \\
& & EEG+EOG & \cellcolor{green!25}0.3521$\pm$0.0048 & \cellcolor{green!25}0.2375$\pm$0.0038 & \cellcolor{green!25}0.3494$\pm$0.0019 \\
& & EEG+ECG & \cellcolor{green!25}\underline{0.3558}$\pm$0.0075 & \cellcolor{green!25}\underline{0.2431}$\pm$0.0052 & \cellcolor{green!25}\underline{0.3550}$\pm$0.0031 \\
& & EOG+ECG & \cellcolor{green!25}0.2587$\pm$0.0053 & \cellcolor{green!25}0.1417$\pm$0.0074 & \cellcolor{green!25}0.2744$\pm$0.0071 \\
& & EEG+EOG+ECG & \cellcolor{green!25}\textbf{0.3642}$\pm$0.0065 & \cellcolor{green!25}\textbf{0.2539}$\pm$0.0041 & \cellcolor{green!25}\textbf{0.3647}$\pm$0.0025 \\
\bottomrule
\end{tabular}
}
\label{tab:seed7}
\end{table}

\begin{table}[H]
\centering
\caption{The results of different methods on HMC.}
\resizebox{\textwidth}{!}{
\begin{tabular}{@{}lccccc@{}}
\toprule
\textbf{Method} & \textbf{Training Modality} & \textbf{Test Modality} & \textbf{Balanced Accuracy} & \textbf{Cohen's Kappa} & \textbf{Weighted F1} \\
\midrule
EEG-Conformer \cite{9991178} & EEG & EEG & \cellcolor{blue!25}0.6767$\pm$0.0200 & \cellcolor{blue!25}0.5886$\pm$0.0397 & \cellcolor{blue!25}0.6550$\pm$0.0463 \\
\multirow{3}*{BIOT \cite{NEURIPS2023_f6b30f3e}} & EEG & EEG & \cellcolor{blue!25}0.6862$\pm$0.0041 & \cellcolor{blue!25}0.6295$\pm$0.0113 & \cellcolor{blue!25}0.7091$\pm$0.0147 \\
& EOG & EOG & \cellcolor{yellow!25}\textbf{0.6192}$\pm$0.0049 & \cellcolor{yellow!25}\underline{0.5553}$\pm$0.0020 & \cellcolor{yellow!25}\textbf{0.6595}$\pm$0.0038 \\
& EMG & EMG & \cellcolor{gray!25}0.3705$\pm$0.0117 & \cellcolor{gray!25}\underline{0.2072}$\pm$0.0050 & \cellcolor{gray!25}\underline{0.3946}$\pm$0.0065 \\
LaBraM-Base \cite{jiang2024large} & EEG & EEG & \cellcolor{blue!25}\underline{0.7286}$\pm$0.0101 & \cellcolor{blue!25}\underline{0.6812}$\pm$0.0073 & \cellcolor{blue!25}\underline{0.7554}$\pm$0.0024 \\
CBraMod \cite{wang2025cbramod} & EEG & EEG & \cellcolor{blue!25}0.7177$\pm$0.0072 & \cellcolor{blue!25}0.6653$\pm$0.0057 & \cellcolor{blue!25}0.7388$\pm$0.0052 \\
\multirow{2}*{Fu \textit{et al.} \cite{10340491}} & EEG+EOG & EOG & \cellcolor{yellow!25}0.4885$\pm$0.0704 & \cellcolor{yellow!25}0.3333$\pm$0.1251 & \cellcolor{yellow!25}0.3754$\pm$0.1670 \\
& EEG+EMG & EMG & \cellcolor{gray!25}\textbf{0.3987}$\pm$0.0057 & \cellcolor{gray!25}0.1937$\pm$0.0052 & \cellcolor{gray!25}0.228$\pm$0.0019 \\
SleepMG \cite{10.1145/3664647.3680854} & EEG+EOG+EMG & EEG+EOG+EMG & \cellcolor{green!25}0.6924$\pm$0.0091 & \cellcolor{green!25}0.6328$\pm$0.0102 & \cellcolor{green!25}0.7168$\pm$0.0073 \\
FeatFusion & EEG+EOG+EMG & EEG+EOG+EMG & \cellcolor{green!25}\textbf{0.7478}$\pm$0.0038 & \cellcolor{green!25}0.6981$\pm$0.0004 & \cellcolor{green!25}0.7728$\pm$0.0010 \\
\midrule
\multirow{7}*{\method} & \multirow{7}*{EEG+EOG+EMG} & EEG & \cellcolor{blue!25}\textbf{0.7289}$\pm$0.0010 & \cellcolor{blue!25}\textbf{0.6880}$\pm$0.0097 & \cellcolor{blue!25}\textbf{0.7635}$\pm$0.0053 \\
& & EOG & \cellcolor{yellow!25}\underline{0.6066}$\pm$0.0073 & \cellcolor{yellow!25}\textbf{0.5554}$\pm$0.0023 & \cellcolor{yellow!25}\underline{0.6533}$\pm$0.0026 \\
& & EMG & \cellcolor{gray!25}\underline{0.3914}$\pm$0.0113 & \cellcolor{gray!25}\textbf{0.2454}$\pm$0.0095 & \cellcolor{gray!25}\textbf{0.4104}$\pm$0.0108 \\
& & EEG+EOG & \cellcolor{green!25}\underline{0.7404}$\pm$0.0018 & \cellcolor{green!25}\underline{0.7063}$\pm$0.0105 & \cellcolor{green!25}\underline{0.7755}$\pm$0.0058 \\
& & EEG+EMG & \cellcolor{green!25}0.7300$\pm$0.0062 & \cellcolor{green!25}0.6958$\pm$0.0070 & \cellcolor{green!25}0.7680$\pm$0.0028 \\
& & EOG+EMG & \cellcolor{green!25}0.6026$\pm$0.0038 & \cellcolor{green!25}0.5717$\pm$0.0108 & \cellcolor{green!25}0.6602$\pm$0.0082 \\
& & EEG+EOG+EMG & \cellcolor{green!25}0.7377$\pm$0.0056 & \cellcolor{green!25}\textbf{0.7120}$\pm$0.0085 & \cellcolor{green!25}\textbf{0.7779}$\pm$0.0031 \\
\bottomrule
\end{tabular}
}
\label{tab:hmc}
\end{table}

\begin{table}[H]
\centering
\caption{The results of different methods on FBM.}
\resizebox{\textwidth}{!}{
\begin{tabular}{@{}lccccc@{}}
\toprule
\textbf{Method} & \textbf{Training Modality} & \textbf{Test Modality} & \textbf{RMSE$\downarrow$} & \textbf{Pearson Correlation} & \textbf{R$^2$ Score} \\
\midrule
EEG-Conformer \cite{9991178} & EEG & EEG & \cellcolor{blue!25}5.7243$\pm$0.4059 & \cellcolor{blue!25}0.4996$\pm$0.0824 & \cellcolor{blue!25}0.2053$\pm$0.1120 \\
\multirow{3}*{BIOT \cite{NEURIPS2023_f6b30f3e}} & EEG & EEG & \cellcolor{blue!25}6.0831$\pm$0.0606 & \cellcolor{blue!25}0.4614$\pm$0.0056 & \cellcolor{blue!25}-0.1714$\pm$0.0771 \\
& EOG & EOG & \cellcolor{yellow!25}6.4821$\pm$0.0357 & \cellcolor{yellow!25}\underline{0.3469}$\pm$0.0079 & \cellcolor{yellow!25}-0.1506$\pm$0.0370 \\
& EMG & EMG & \cellcolor{gray!25}6.1194$\pm$0.3700 & \cellcolor{gray!25}0.3950$\pm$0.1642 & \cellcolor{gray!25}-0.2671$\pm$0.0753 \\
LaBraM-Base \cite{jiang2024large} & EEG & EEG & \cellcolor{blue!25}\underline{5.0466}$\pm$0.0112 & \cellcolor{blue!25}\textbf{0.6567}$\pm$0.0020 & \cellcolor{blue!25}\textbf{0.3668}$\pm$0.0043 \\
CBraMod \cite{wang2025cbramod} & EEG & EEG & \cellcolor{blue!25}5.1216$\pm$0.0174 & \cellcolor{blue!25}\underline{0.6345}$\pm$0.0035 & \cellcolor{blue!25}\underline{0.3490}$\pm$0.0033 \\
\multirow{2}*{Fu \textit{et al.} \cite{10340491}} & EEG+EMG & EEG & \cellcolor{blue!25}5.4682$\pm$0.0457 & \cellcolor{blue!25}0.5391$\pm$0.0148 & \cellcolor{blue!25}0.2721$\pm$0.0128 \\
& EOG+EMG & EOG & \cellcolor{yellow!25}\underline{6.2401}$\pm$0.0722 & \cellcolor{yellow!25}0.3278$\pm$0.0323 & \cellcolor{yellow!25}\textbf{0.0888}$\pm$0.0227 \\
FeatFusion & EEG+EOG+EMG & EEG+EOG+EMG & \cellcolor{green!25}5.0511$\pm$0.1564 & \cellcolor{green!25}0.6443$\pm$0.0203 & \cellcolor{green!25}0.2674$\pm$0.0456 \\
\midrule
\multirow{7}*{\method} & \multirow{7}*{EEG+EOG+EMG} & EEG & \cellcolor{blue!25}\textbf{4.9650}$\pm$0.0089 & \cellcolor{blue!25}0.6313$\pm$0.0075 & \cellcolor{blue!25}0.3122$\pm$0.0068 \\
& & EOG & \cellcolor{yellow!25}\textbf{6.0321}$\pm$0.0217 & \cellcolor{yellow!25}\textbf{0.3838}$\pm$0.0042 & \cellcolor{yellow!25}\underline{0.0173}$\pm$0.0156 \\
& & EMG & \cellcolor{gray!25}\textbf{5.5950}$\pm$0.1048 & \cellcolor{gray!25}\textbf{0.5934}$\pm$0.0122 & \cellcolor{gray!25}\textbf{0.0337}$\pm$0.0226 \\
& & EEG+EOG & \cellcolor{green!25}4.9002$\pm$0.0312 & \cellcolor{green!25}0.6126$\pm$0.0059 & \cellcolor{green!25}0.3472$\pm$0.0079 \\
& & EEG+EMG & \cellcolor{green!25}\underline{4.7133}$\pm$0.0354 & \cellcolor{green!25}\textbf{0.6637}$\pm$0.0023 & \cellcolor{green!25}\underline{0.3613}$\pm$0.0111 \\
& & EOG+EMG & \cellcolor{green!25}5.0091$\pm$0.0184 & \cellcolor{green!25}0.5979$\pm$0.0060 & \cellcolor{green!25}0.2812$\pm$0.0038 \\
& & EEG+EOG+EMG & \cellcolor{green!25}\textbf{4.6191}$\pm$0.0122 & \cellcolor{green!25}\underline{0.6580}$\pm$0.0008 & \cellcolor{green!25}\textbf{0.3995}$\pm$0.0074 \\
\bottomrule
\end{tabular}
}
\label{tab:fbm}
\end{table}

\subsection{Experimental Results}
The results are presented in Table~\ref{tab:seed7}, \ref{tab:hmc}, and \ref{tab:fbm}, with baseline descriptions provided in Appendix~\ref{baseline}. Overall, \method achieves competitive performance compared to a wide range of unimodal, multimodal, and cross-modal methods. Notably, there is a trade-off in selecting the best model for \method, as different modality combinations reach optimal performance at different training epochs, placing our approach at a disadvantage. Despite this, \method matches or surpasses the performance of the leading baselines in most cases. In addition to achieving best results under the single EEG condition, \method also outperforms all baselines on other unimodal (EOG, ECG, EMG) and multimodal settings. Specifically, it delivers superior performance for EOG, ECG, and multimodal scenarios, and remains competitive with LaBraM. On HMC, \method consistently achieves the best results across all modalities except EOG. While it does not outperform the top baselines in terms of correlation and $R^2$, it achieves the lowest RMSE on the FBM dataset.

\subsection{Ablation Study}
\begin{figure}[t!]
    \centering
    \includegraphics[width=1\linewidth]{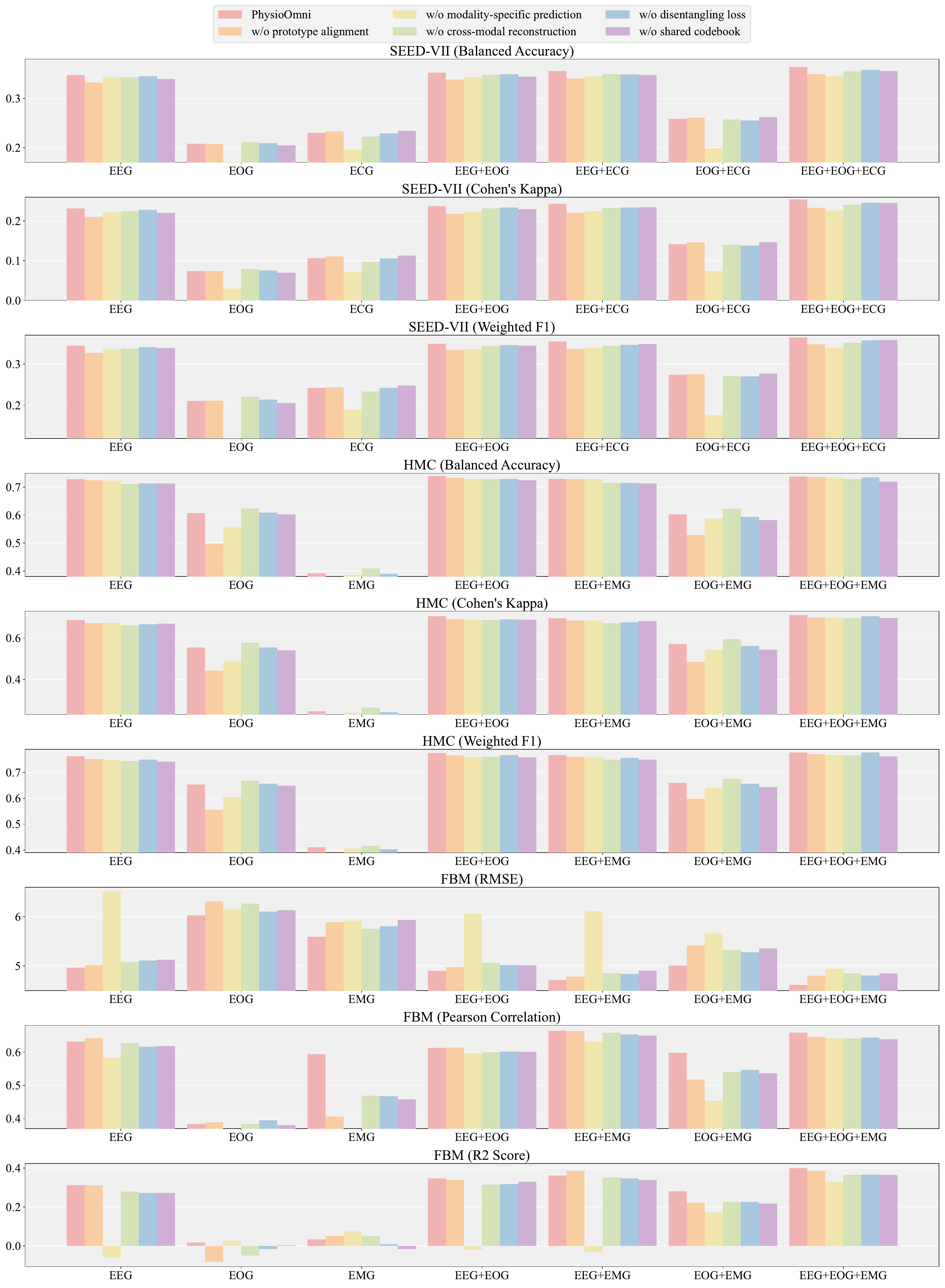}
    \vspace{-15pt}
    \caption{Ablation study of pivotal components on downstream datasets.}
    \label{ablation}
\end{figure}

To evaluate the contribution of key components, we conduct ablation studies on modality-specific prediction, disentangling loss, prototype alignment, cross-modal reconstruction, and the shared codebook. The results are presented in Figure~\ref{ablation}, where lower RMSE values on FBM indicate better performance. Overall, the removal of any component generally leads to performance degradation. Notably, prototype alignment proves especially effective on SEED-VII and HMC, while modality-specific prediction is crucial across all datasets. Removing cross-modal reconstruction, disentangling loss, or the shared codebook consistently results in performance drops, underscoring their importance.

\section{Conclusion}
We propose \method, a universal multimodal physiological foundation model that learns robust representations through decoupled tokenization and masked signal modeling, and effectively handles arbitrary missing modalities via resilient fine-tuning. Through comprehensive experiments on four BCI tasks, \method consistently achieves SOTA across both unimodal and multimodal settings. 



\bibliography{reference}
\bibliographystyle{plain}


\appendix

\appendix
\section{Hyperparameter Settings}
\label{hyperparameters}

\begin{table}[H]
\centering
\caption{Hyperparameters for decoupled multimodal tokenizer.}
\begin{tabular}{@{}cccc@{}}
\toprule
\multicolumn{2}{c}{\textbf{Hyperparameters}} & \textbf{EEG Encoder} & \textbf{Other Encoders} \cr
\midrule
\multirow{5}*{Temporal Encoder} & Iput channels & \multicolumn{2}{c}{\{1,8,8\}} \\
& Output channels & \multicolumn{2}{c}{\{16,16,16\}} \\
& Kernel size & \multicolumn{2}{c}{\{15,3,3\}} \\
& Stride & \multicolumn{2}{c}{\{8,1,1\}} \\
& Padding & \multicolumn{2}{c}{\{7,1,1\}} \\
\midrule
\multicolumn{2}{c}{Transformer encoder layers} & \multicolumn{2}{c}{12} \\
\multicolumn{2}{c}{Transformer decoder layers} & \multicolumn{2}{c}{3} \\
\multicolumn{2}{c}{Hidden size} & 200 & 100 \\
\multicolumn{2}{c}{MLP size} & 800 & 400 \\
\multicolumn{2}{c}{Attention head number} & \multicolumn{2}{c}{10} \\
\multicolumn{2}{c}{Shared Codebook size} & \multicolumn{2}{c}{8192$\times$64} \\
\multicolumn{2}{c}{Private Codebook size} & \multicolumn{2}{c}{8192$\times$64} \\
\midrule
\multicolumn{2}{c}{Batch size} & \multicolumn{2}{c}{512} \\
\multicolumn{2}{c}{Peak learning rate} & \multicolumn{2}{c}{1e-4} \\
\multicolumn{2}{c}{Minimal learning rate} & \multicolumn{2}{c}{1e-5} \\
\multicolumn{2}{c}{Learning rate scheduler} & \multicolumn{2}{c}{Cosine} \\
\multicolumn{2}{c}{Optimizer} & \multicolumn{2}{c}{AdamW} \\
\multicolumn{2}{c}{Adam $\beta$} & \multicolumn{2}{c}{(0.9,0.99)} \\
\multicolumn{2}{c}{Weight decay} & \multicolumn{2}{c}{1e-4} \\
\multicolumn{2}{c}{$\alpha_1,\alpha_2$} & \multicolumn{2}{c}{(1,0.1)} \\
\multicolumn{2}{c}{Total epochs} & \multicolumn{2}{c}{100} \\
\multicolumn{2}{c}{Warmup epochs} & \multicolumn{2}{c}{10} \\
\multicolumn{2}{c}{Data overlap} & \multicolumn{2}{c}{None} \\
\multicolumn{2}{c}{Gradient clipping} & \multicolumn{2}{c}{None} \\
\bottomrule
\end{tabular}
\label{tab:vq}
\end{table}

\begin{table}[H]
\centering
\caption{Hyperparameters for masked signal pre-training.}
\begin{tabular}{@{}cccc@{}}
\toprule
\multicolumn{2}{c}{\textbf{Hyperparameters}} & \textbf{EEG Encoder}  & \textbf{Other Encoders} \cr
\midrule
\multirow{5}*{Temporal Encoder} & Iput channels & \multicolumn{2}{c}{\{1,8,8\}} \\
& Output channels & \multicolumn{2}{c}{\{16,16,16\}} \\
& Kernel size & \multicolumn{2}{c}{\{15,3,3\}} \\
& Stride & \multicolumn{2}{c}{\{8,1,1\}} \\
& Padding & \multicolumn{2}{c}{\{7,1,1\}} \\
\midrule
\multicolumn{2}{c}{Transformer encoder layers} & \multicolumn{2}{c}{12} \\
\multicolumn{2}{c}{Hidden size} & 200 & 100 \\
\multicolumn{2}{c}{MLP size} & 800 & 400 \\
\multicolumn{2}{c}{Attention head number} & \multicolumn{2}{c}{10} \\
\midrule
\multicolumn{2}{c}{Mask ratio} & \multicolumn{2}{c}{0.5 (EEG, EMG), 0.7 (EOG, ECG)} \\
\multicolumn{2}{c}{Batch size} & \multicolumn{2}{c}{512} \\
\multicolumn{2}{c}{Peak learning rate} & \multicolumn{2}{c}{5e-4}\\
\multicolumn{2}{c}{Minimal learning rate} & \multicolumn{2}{c}{1e-5} \\
\multicolumn{2}{c}{Learning rate scheduler} & \multicolumn{2}{c}{Cosine} \\
\multicolumn{2}{c}{Optimizer} & \multicolumn{2}{c}{AdamW} \\
\multicolumn{2}{c}{Adam $\beta$} & \multicolumn{2}{c}{(0.9,0.98)} \\
\multicolumn{2}{c}{Weight decay} & \multicolumn{2}{c}{0.05} \\
\multicolumn{2}{c}{Total epochs} & \multicolumn{2}{c}{50} \\
\multicolumn{2}{c}{Warmup epochs} & \multicolumn{2}{c}{5} \\
\multicolumn{2}{c}{Data overlap} & \multicolumn{2}{c}{None} \\
\multicolumn{2}{c}{Gradient clipping} & \multicolumn{2}{c}{3} \\
\bottomrule
\end{tabular}
\label{tab:pretraining}
\end{table}

\begin{table}[H]
\centering
\caption{Hyperparameters for fine-tuning.}
\begin{tabular}{@{}cc@{}}
\toprule
\textbf{Hyperparameters} & \textbf{Values} \cr
\midrule
Projected feature size & 128$\times$128 \\
MoE layers & 1 \\
Expert number & 4 \\
Hidden size & 128 \\
MLP size & 512 \\
Attention head number & 10 \\
Number of prototypes & 7 (SEED-VII), 5 (HMC), 256 (FBM), 2 (EEGMAT) \\
\midrule
Batch size & 128 (EEGMAT), 512 (SEED-VII, HMC, FBM) \\
\multirow{4}*{Loss ratio $\gamma$} & SEED-VII: 1 (align), 0.5 (main), 0.5 (EEG), 0.5 (EOG), 0.5 (ECG) \\
& HMC: 1 (align), 0.1 (main), 0.01 (EEG), 4 (EOG), 0.5 (EMG) \\
& FBM: 0.01 (align), 0.1 (main), 2 (EEG), 0.5 (EOG), 1 (EMG) \\
& EEGMAT: 1 (align), 0.5 (main), 0.5 (EEG), 0.5 (EOG), 0.5 (EMG) \\
Peak learning rate & 1e-3 (SEED-VII), 5e-4 (HMC, FBM), 1e-4 (EEGMAT) \\
Minimal learning rate & 1e-4 (SEED-VII), 5e-5 (HMC, FBM), 1e-5 (EEGMAT) \\
Learning rate scheduler & Cosine \\
Optimizer & AdamW \\
Adam $\beta$ & (0.9,0.999) \\
Weight decay & 0.05 \\
Total epochs & 50 \\
Warmup epochs & 5 \\
Gradient clipping & None \\
Label smoothing & 0.1 (multi-class classification) \\
\bottomrule
\end{tabular}
\label{tab:instr}
\end{table}

\section{Pre-training Datasets}
\label{data}
We utilize a diverse collection of multimodal physiological datasets for various tasks:

\begin{itemize}
    \item \textbf{TUEG} \cite{obeid2016temple}: The TUH EEG Corpus (TUEG) is an extensive archive comprising 26,846 clinical EEG recordings collected at Temple University Hospital. Some recordings include EOG (horizontal), ECG, and EMG signals alongside EEG (21–23 channels), with sampling frequencies ranging from 250 to 1024 Hz. For pre-training, we select recordings that include at least three modalities.
    \item \textbf{DEAP} \cite{5871728}: DEAP provides a multimodal dataset for analyzing human affective states, recorded from 32 participants as they watched 40 one-minute music video excerpts. EEG (32 channels), EOG (horizontal and vertical), and EMG signals were captured at a sampling rate of 512 Hz.
    \item \textbf{Sleep-EDF} \cite{867928}: This dataset contains 197 whole-night sleep recordings featuring EEG (2 channels), EOG (horizontal), and chin EMG signals, sampled at 100 Hz. Hypnograms, representing sleep patterns, were manually scored by trained technicians following the Rechtschaffen and Kales manual.
    \item \textbf{CAP} \cite{terzano2005cyclic}: The CAP Sleep Database includes 108 polysomnographic recordings sampled at 128 Hz, featuring EEG (3 or more channels), EOG (horizontal and vertical), and EMG signals, with annotations for sleep stages and Cyclic Alternating Pattern (CAP).
    \item \textbf{GX} \cite{gebodh2021dataset}: This dataset combines high-density EEG (30 channels) with ECG and EOG (horizontal) during transcranial electrical stimulation (tES) across 783 trials and 62 sessions, sampled at 2000 Hz. It includes nine HD-tES types targeting three cortical regions with different waveforms, with participants performing vigilance tasks, completing wellness questionnaires, and undergoing repeated sessions to assess within-participant reliability.
    \item \textbf{Private data}: This dataset comprises 54 recordings from 19 subjects while watching videos, with each recording lasting approximately one hour. EEG (62 channels), EOG (horizontal and vertical), and ECG signals were captured using the ESI NeuroScan System at a sampling rate of 1000 Hz.
\end{itemize}

\section{Baselines}
\label{baseline}
To comprehensively evaluate \method, we compare it with the following classical and state-of-the-art baselines:
\begin{itemize}
    \item \textbf{EEG-Conformer} \cite{9991178}: A compact Convolutional Transformer that integrates local feature extraction via convolution and global feature modeling via self-attention for EEG classification. By combining temporal and spatial convolutions with self-attention, EEG-Conformer effectively captures both short-term and long-term dependencies in EEG signals.  
    \item \textbf{BIOT} \cite{NEURIPS2023_f6b30f3e}: A Biosignal Transformer designed for flexible biosignal encoding, BIOT facilitates cross-data learning across diverse signal formats such as EEG and ECG. It tokenizes each channel into fixed-length segments while preserving spatio-temporal features through channel and positional embeddings, demonstrating strong generalizability via joint pre-training and fine-tuning on multiple biosignal tasks. 
    \item \textbf{LaBraM} \cite{jiang2024large}: A unified foundation model for EEG, LaBraM enables cross-dataset learning by segmenting EEG signals into channel patches and encoding them with a vector-quantized neural tokenizer. It employs masked neural code prediction for unsupervised pre-training, leveraging approximately 2,500 hours of EEG data from 20 datasets. We fine-tune its publicly available pre-trained checkpoints on each downstream dataset.  
    \item \textbf{CBraMod} \cite{wang2025cbramod}: An EEG foundation model designed to address the heterogeneity of spatial and temporal dependencies in EEG signals, CBraMod utilizes a criss-cross Transformer with separate attention mechanisms. It incorporates an asymmetric conditional positional encoding scheme for enhanced adaptability across diverse EEG formats. We fine-tune its publicly available pre-trained checkpoints on each downstream dataset.  
    \item \textbf{Fu \textit{et al.}} \cite{10340491}: This method employs a multimodal training strategy using supervised contrastive learning, leveraging EMG signals during training to enhance EEG-based gait classification and regression. The model learns gait patterns from EEG with EMG guidance but relies solely on EEG during inference. We use multimodal signals to train and single modality to test for this method.
    \item \textbf{SleepMG} \cite{10.1145/3664647.3680854}: A multimodal generalizable sleep staging method, SleepMG balances inter-modal differences in PSG by assessing classification and domain discrimination performances across modalities. It defines modal performance metrics based on variance from the average performance and adaptively adjusts gradient updates to emphasize poorly balanced modalities.  
    \item \textbf{FeatFusion}: We leverage our pre-trained encoders from \method and fuse the features extracted from these encoders by the Homogeneous Representation Mapping and Feature Fuser. We concatenate the features from all modalities after Homogeneous Representation Mapping and feed them into the Transformer layer. This approach utilizes the generic representations of each modality and employs proposed fusion strategy.
\end{itemize}

\section{Evaluation Metrics}
We consider the following metrics to comprehensively evaluate all methods:
\begin{itemize}
    \item \textbf{Balanced Accuracy}: This metric calculates the average of recall (true positive rate) for each class, ensuring that each class contributes equally to the final accuracy score. It is particularly useful when dealing with imbalanced datasets.
    \item \textbf{AUC-PR}: AUC-PR is the area under the precision-recall curve, which plots precision against recall at various thresholds. It is especially useful for imbalanced datasets where the negative class is overwhelming.
    \item \textbf{AUROC}: AUROC is the area under the ROC curve, which plots the true positive rate (sensitivity) against the false positive rate (1-specificity) at various thresholds. It provides a comprehensive view of a model’s performance across all classification thresholds.
    \item \textbf{Cohen’s Kappa}: Cohen's Kappa measures the agreement between two raters, correcting for chance agreement. A Kappa value of 1 indicates perfect agreement, while a value of 0 indicates no agreement beyond chance, with negative values indicating worse than random agreement.
    \item \textbf{Weighted F1}: The F1 score is the harmonic mean of precision and recall, balancing the trade-off between false positives and false negatives. The weighted F1 takes into account the support (the number of true instances) of each class, so it gives more importance to classes with more data.
    \item \textbf{RMSE}: RMSE is the square root of the average squared differences between predicted and actual values, providing a measure of the model's prediction error. Larger errors are penalized more due to the squaring of differences while a lower RMSE indicates a better fit between the model and the data.
    \item \textbf{R$^2$ Score}: R$^2$ measures the proportion of the variance in the dependent variable that is predictable from the independent variables. A value of 1 indicates perfect prediction, 0 indicates that the model does not improve on simply predicting the mean of the data, and negative values suggest that the model is performing worse than a simple mean predictor.
    \item \textbf{Pearson Correlation}: Pearson correlation quantifies the linear relationship between two variables, with values ranging from -1 (perfect negative correlation) to +1 (perfect positive correlation). A value of 0 indicates no linear relationship.
\end{itemize}
For the monitor metric, AUROC is used for binary classification, Cohen's Kappa for multi-class classification, and R$^2$ Score for regression.

\section{More Experiments on EEGMAT}
\label{EEGMAT}
This dataset \cite{zyma2019electroencephalograms} supports the investigation of EEG characteristics during mental tasks, including Fourier power spectral, coherence, and detrended fluctuation analysis. It contains EEG recordings from 36 healthy volunteers performing a mental subtraction task, categorized into good and bad counters, and is available through the Physiobank platform for neuroscience research on cognitive workload. Subjects from number 0 to 25 are used for training, 26 to 30 for validation, and 31 to 35 for testing.

The results are presented in Table~\ref{tab:eegmat}. Across all evaluation metrics and modality configurations, \method consistently outperforms all competing baselines, demonstrating its strong generalization and adaptability. When trained and tested on both EEG and ECG modalities, \method achieves the highest performance. Even in unimodal scenarios, \method maintains its superiority, outperforming dedicated EEG and ECG baselines such as BIOT, LaBraM, and CBraMod. These results suggest that EEG generally provides stronger predictive signals than ECG for this task, yet combining both modalities leads to further performance improvements. This highlights the effectiveness of \method in leveraging complementary information from multiple physiological signals, both in single-modality and multimodal settings.

\begin{table}[H]
\centering
\caption{The results of different methods on EEGMAT.}
\resizebox{\textwidth}{!}{
\begin{tabular}{@{}lccccc@{}}
\toprule
\textbf{Method} & \textbf{Training Modality} & \textbf{Test Modality} & \textbf{Balanced Accuracy} & \textbf{AUC-PR} & \textbf{AUROC} \\
\midrule
EEG-Conformer \cite{9991178} & EEG & EEG & \cellcolor{blue!25}0.5141$\pm$0.0060 & \cellcolor{blue!25}0.6417$\pm$0.0386 & \cellcolor{blue!25}0.6455$\pm$0.0227 \\
\multirow{2}*{BIOT \cite{NEURIPS2023_f6b30f3e}} & EEG & EEG & \cellcolor{blue!25}\underline{0.6655}$\pm$0.0665 & \cellcolor{blue!25}\underline{0.7189}$\pm$0.0722 & \cellcolor{blue!25}\underline{0.7342}$\pm$0.0536 \\
& ECG & ECG & \cellcolor{red!25}0.5391$\pm$0.0043 & \cellcolor{red!25}0.5974$\pm$0.0293 & \cellcolor{red!25}0.5656$\pm$0.0263 \\
LaBraM-Base \cite{jiang2024large} & EEG & EEG & \cellcolor{blue!25}0.6609$\pm$0.0204 & \cellcolor{blue!25}0.7174$\pm$0.0234 & \cellcolor{blue!25}0.7272$\pm$0.0165 \\
CBraMod \cite{wang2025cbramod} & EEG & EEG & \cellcolor{blue!25}0.6310$\pm$0.0129 & \cellcolor{blue!25}0.7073$\pm$0.0322 & \cellcolor{blue!25}0.7303$\pm$0.0225 \\
\multirow{2}*{Fu \textit{et al.} \cite{10340491}} & EEG+ECG & EEG & \cellcolor{blue!25}0.5241 $\pm$ 0.0734 & \cellcolor{blue!25}0.5513 $\pm$ 0.0784 & \cellcolor{blue!25}0.5737 $\pm$ 0.0621\\
& EEG+ECG & ECG & \cellcolor{red!25}\underline{0.5586}$\pm$ 0.0049 & \cellcolor{red!25}\underline{0.6042}$\pm$ 0.0249 & \cellcolor{red!25}\underline{0.5767}$\pm$ 0.0548 \\
FeatFusion & EEG+ECG & EEG+ECG & \cellcolor{green!25}0.7219$\pm$0.0809 & \cellcolor{green!25}\textbf{0.9164}$\pm$0.0131 & \cellcolor{green!25}0.8873$\pm$0.0322 \\
\midrule
\multirow{3}*{\method} & \multirow{3}*{EEG+ECG} & EEG & \cellcolor{blue!25}\textbf{0.7637}$\pm$0.0282 & \cellcolor{blue!25}\textbf{0.8410}$\pm$0.0135 & \cellcolor{blue!25}\textbf{0.8036}$\pm$0.0216 \\
& & ECG & \cellcolor{red!25}\textbf{0.7471}$\pm$0.0496 & \cellcolor{red!25}\textbf{0.7718}$\pm$0.0798 & \cellcolor{red!25}\textbf{0.8173}$\pm$0.0569 \\
& & EEG+ECG & \cellcolor{green!25}\textbf{0.7870}$\pm$0.0270 & \cellcolor{green!25}0.9001$\pm$0.0261 & \cellcolor{green!25}\textbf{0.8912}$\pm$0.0257 \\
\bottomrule
\end{tabular}
}
\label{tab:eegmat}
\end{table}

\section{Ablation on Frozen Encoders}
\begin{figure}[H]
    \centering
    \includegraphics[width=1\linewidth]{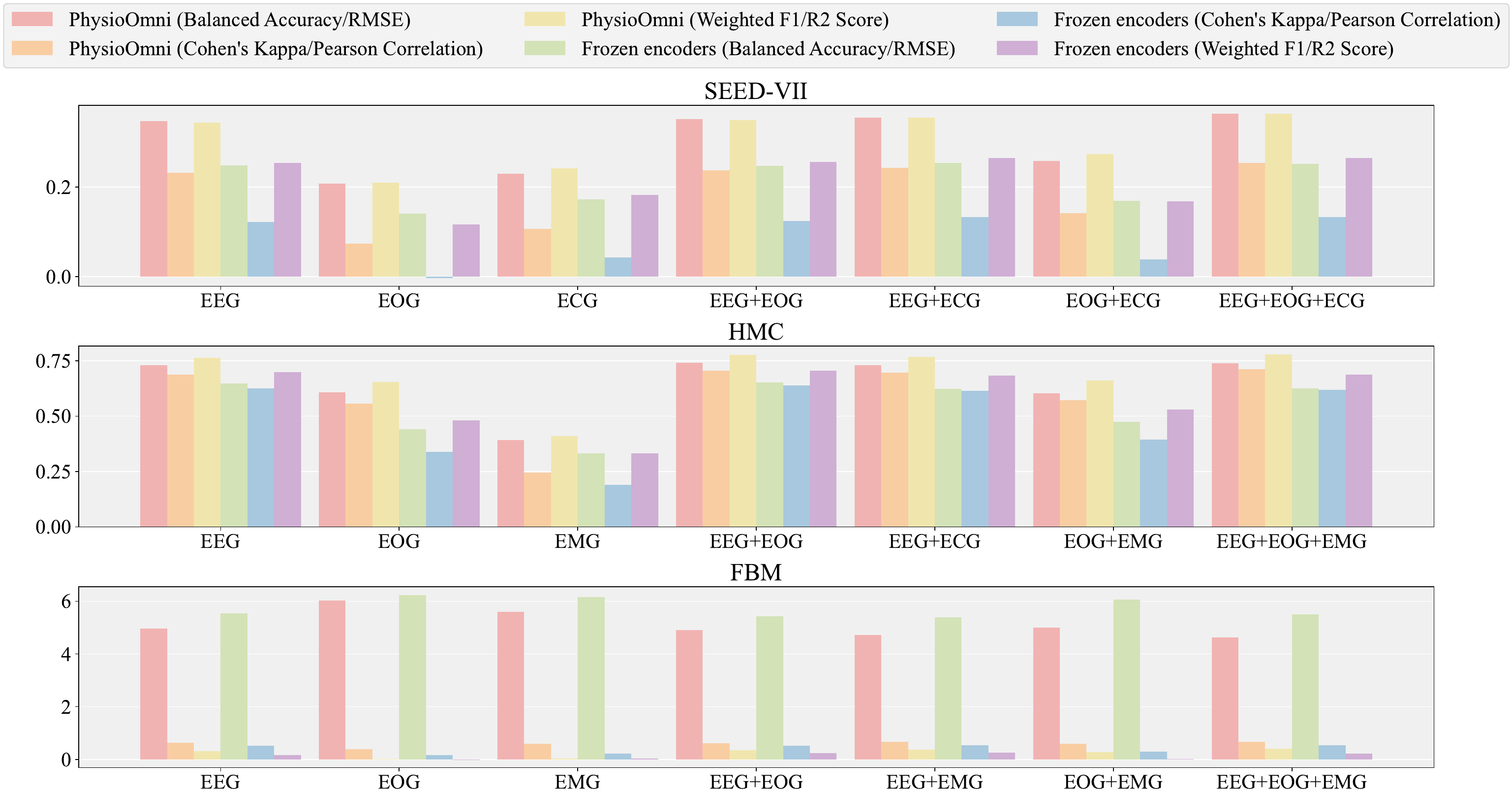}
    \vspace{-15pt}
    \caption{Ablation study on frozen encoders.}
    \label{frozen}
\end{figure}

We further investigate the impact of freezing the pre-trained encoders during fine-tuning, with results presented in Figure~\ref{frozen}. As expected, freezing the encoders leads to a significant drop in downstream performance across all tasks and modality combinations. This highlights the importance of full end-to-end fine-tuning, which allows the model to adapt pre-trained representations to task-specific nuances and optimize the fusion of multimodal features. The results emphasize that while pre-training provides a strong initialization, fine-tuning remains crucial for maximizing performance in real-world downstream applications.

\section{Ablation on Pre-training}
\begin{figure}[b]
    \centering
    \includegraphics[width=1\linewidth]{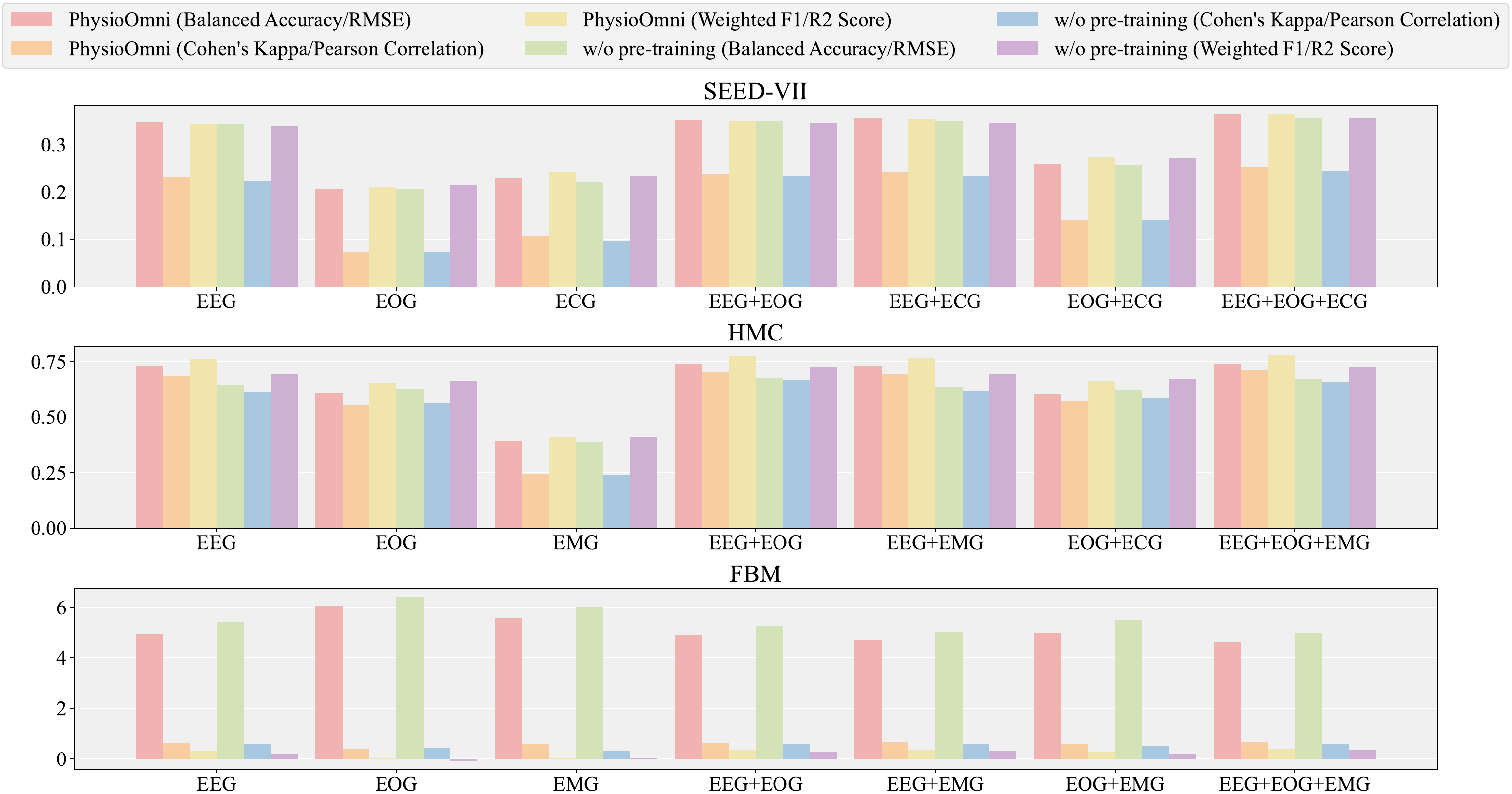}
    \vspace{-15pt}
    \caption{Ablation study on pre-training.}
    \label{pretrain}
\end{figure}
To assess the effectiveness of our proposed pre-training paradigm, we conduct an ablation study by randomly initializing all encoder parameters. The results, shown in Figure~\ref{pretrain}, demonstrate a consistent performance drop across all datasets and modality combinations when pre-training is removed. This highlights the critical role of pre-training in enabling robust and generalizable representations. Without pre-training, the model struggles to extract meaningful features from the raw physiological signals, leading to degraded performance even after fine-tuning. In contrast, the pre-trained encoders provide a strong initialization, capturing essential modality-specific and modality-invariant patterns that transfer well across downstream tasks. These findings confirm that our masked signal modeling and codebook-based design are effective in learning transferable representations that significantly enhance overall model performance.

\section{Ablation on Homogeneous Representation Mapping \& Feature Fuser}
Homogeneous Representation Mapping (HRM) is designed to project representations from different modalities into a common space with uniform embedding sizes, ensuring better compatibility across modalities. Meanwhile, the Feature Fuser (FF) integrates these projected representations to generate comprehensive multimodal features, enhancing the model's ability to leverage complementary information across different signals. 

To assess the contribution of HRM and FF, we conduct an ablation study and present the results in Table~\ref{tab:hrm}. The findings demonstrate that removing the two modules generally lead to a degradation in performance on SEED-VII and HMC, underscoring their crucial role in effective modality fusion. However, on FBM, while eliminating HRM and FF leads to significant performance degration for EEG, it results in a performance boost in certain cases, particularly for EEG in combination with another modality. A possible explanation is that FBM exhibits a different modality interaction pattern compared to SEED-VII and HMC, where the added complexity of fusion modules may lead to overfitting. These results suggest that for FBM, a simpler fusion mechanism could be more advantageous than enforcing strict homogeneous representation constraints. This highlights the importance of dataset-specific tuning when designing multimodal fusion strategies.

\begin{table}[H]
\centering
\caption{The ablation study on Homogeneous Representation Mapping \& Feature Fuser (\method\ /\ w/o HRM \& FF).}
\resizebox{\textwidth}{!}{
\begin{tabular}{@{}cccc@{}}
\toprule
\multirow{2}*{\textbf{Test Modality}} & \multicolumn{3}{c}{\textbf{SEED-VII}} \\
\cmidrule(l){2-4}
 & \textbf{Balanced Accuracy} & \textbf{Cohen's Kappa} & \textbf{Weighted F1} \\
\midrule
EEG & 0.3479$\pm$0.0054 / \textcolor{blue}{0.3444$\pm$0.0008} & 0.2316$\pm$0.0045 / \textcolor{blue}{0.2295$\pm$0.0011} & 0.3442$\pm$0.0025 / \textcolor{blue}{0.3444$\pm$0.0006} \\
EOG & 0.2079$\pm$0.0052 / \textcolor{blue}{0.1841$\pm$0.0243} & 0.0737$\pm$0.0051 / \textcolor{blue}{0.0476$\pm$0.0263} & 0.2104$\pm$0.0084 / \textcolor{blue}{0.1684$\pm$0.0459} \\
ECG & 0.2302$\pm$0.0049 / \textcolor{blue}{0.2279$\pm$0.0020} & 0.1062$\pm$0.0065 / \textcolor{blue}{0.1068$\pm$0.0043} & 0.2425$\pm$0.0058 / \textcolor{blue}{0.2435$\pm$0.0047} \\
EEG+EOG & 0.3521$\pm$0.0048 / \textcolor{blue}{0.3461$\pm$0.0023} & 0.2375$\pm$0.0038 / \textcolor{blue}{0.2326$\pm$0.0033} & 0.3494$\pm$0.0019 / \textcolor{blue}{0.3475$\pm$0.0020} \\
EEG+ECG & 0.3558$\pm$0.0075 / \textcolor{blue}{0.3502$\pm$0.0014} & 0.2431$\pm$0.0052 / \textcolor{blue}{0.2395$\pm$0.0005} & 0.3550$\pm$0.0031 / \textcolor{blue}{0.3538$\pm$0.0013} \\
EOG+ECG & 0.2587$\pm$0.0053 / \textcolor{blue}{0.2405$\pm$0.0091} & 0.1417$\pm$0.0074 / \textcolor{blue}{0.1235$\pm$0.0093} & 0.2744$\pm$0.0071 / \textcolor{blue}{0.2564$\pm$0.0098} \\
EEG+EOG+ECG & 0.3642$\pm$0.0065 / \textcolor{blue}{0.3540$\pm$0.0033} & 0.2539$\pm$0.0041 / \textcolor{blue}{0.2456$\pm$0.0038} & 0.3647$\pm$0.0025 / \textcolor{blue}{0.3594$\pm$0.0025} \\
\midrule
\multirow{2}*{\textbf{Test Modality}} & \multicolumn{3}{c}{\textbf{HMC}} \\
\cmidrule(l){2-4}
& \textbf{Balanced Accuracy} & \textbf{Cohen's Kappa} & \textbf{Weighted F1} \\
\midrule
EEG & 0.7289$\pm$0.0010 / \textcolor{blue}{0.7115$\pm$0.0064} & 0.6880$\pm$0.0097 / \textcolor{blue}{0.6687$\pm$0.0006} & 0.7635$\pm$0.0053 / \textcolor{blue}{0.7460$\pm$0.0028} \\
EOG & 0.6066$\pm$0.0073 / \textcolor{blue}{0.6247$\pm$0.0096} & 0.5554$\pm$0.0023 / \textcolor{blue}{0.5669$\pm$0.0125} & 0.6533$\pm$0.0026 / \textcolor{blue}{0.6683$\pm$0.0067} \\
EMG & 0.3914$\pm$0.0113 / \textcolor{blue}{0.3934$\pm$0.0082} & 0.2454$\pm$0.0095 / \textcolor{blue}{0.2425$\pm$0.0046} & 0.4104$\pm$0.0108 / \textcolor{blue}{0.4133$\pm$0.0142} \\
EEG+EOG & 0.7404$\pm$0.0018 / \textcolor{blue}{0.7252$\pm$0.0066} & 0.7063$\pm$0.0105 / \textcolor{blue}{0.6886$\pm$0.0031} & 0.7755$\pm$0.0058 / \textcolor{blue}{0.7606$\pm$0.0035} \\
EEG+EMG & 0.7300$\pm$0.0062 / \textcolor{blue}{0.7132$\pm$0.0066} & 0.6958$\pm$0.0070 / \textcolor{blue}{0.6769$\pm$0.0034} & 0.7680$\pm$0.0028 / \textcolor{blue}{0.7513$\pm$0.0024} \\
EOG+EMG & 0.6026$\pm$0.0038 / \textcolor{blue}{0.6170$\pm$0.0145} & 0.5717$\pm$0.0108 / \textcolor{blue}{0.5800$\pm$0.0135} & 0.6602$\pm$0.0082 / \textcolor{blue}{0.6737$\pm$0.0109} \\
EEG+EOG+EMG & 0.7377$\pm$0.0056 / \textcolor{blue}{0.7252$\pm$0.0069} & 0.7120$\pm$0.0085 / \textcolor{blue}{0.6961$\pm$0.0033} & 0.7779$\pm$0.0031 / \textcolor{blue}{0.7651$\pm$0.0033} \\
\midrule
\multirow{2}*{\textbf{Test Modality}} & \multicolumn{3}{c}{\textbf{FBM}} \\
\cmidrule(l){2-4}
& \textbf{RMSE$\downarrow$} & \textbf{Pearson Correlation} & \textbf{R$^2$ Score} \\
\midrule
EEG & 4.9650$\pm$0.0089 / \textcolor{blue}{5.1240$\pm$0.0075} & 0.6313$\pm$0.0075 / \textcolor{blue}{0.6411$\pm$0.0023} & 0.3122$\pm$0.0068 / \textcolor{blue}{0.2683$\pm$0.0065} \\
EOG & 6.0321$\pm$0.0217 / \textcolor{blue}{6.0402$\pm$0.1256} & 0.3838$\pm$0.0042 / \textcolor{blue}{0.3946$\pm$0.0077} & 0.0173$\pm$0.0156 / \textcolor{blue}{0.0266$\pm$0.0465} \\
EMG & 5.5950$\pm$0.1048 / \textcolor{blue}{5.5086$\pm$0.5068} & 0.5934$\pm$0.0122 / \textcolor{blue}{0.4499$\pm$0.1828} & 0.0337$\pm$0.0226 / \textcolor{blue}{0.1791$\pm$0.1108} \\
EEG+EOG & 4.9002$\pm$0.0312 / \textcolor{blue}{4.8483$\pm$0.0576} & 0.6126$\pm$0.0059 / \textcolor{blue}{0.6311$\pm$0.0069} & 0.3472$\pm$0.0079 / \textcolor{blue}{0.3712$\pm$0.0139} \\
EEG+EMG & 4.7133$\pm$0.0354 / \textcolor{blue}{4.6800$\pm$0.1101} & 0.6637$\pm$0.0023 / \textcolor{blue}{0.6676$\pm$0.0092} & 0.3613$\pm$0.0111 / \textcolor{blue}{0.4144$\pm$0.0162} \\
EOG+EMG & 5.0091$\pm$0.0184 / \textcolor{blue}{5.3304$\pm$0.3294} & 0.5979$\pm$0.0060 / \textcolor{blue}{0.5268$\pm$0.0759} & 0.2812$\pm$0.0038 / \textcolor{blue}{0.2438$\pm$0.0069} \\
EEG+EOG+EMG & 4.6191$\pm$0.0122 / \textcolor{blue}{4.7337$\pm$0.1369} & 0.6580$\pm$0.0008 / \textcolor{blue}{0.6593$\pm$0.0064} & 0.3995$\pm$0.0074 / \textcolor{blue}{0.4049$\pm$0.0228} \\
\bottomrule
\end{tabular}
}
\label{tab:hrm}
\end{table}

\section{Limitations}
\label{limitations}
While \method demonstrates strong performance and generalizability across diverse physiological signals and downstream tasks, several limitations remain: 1) Although \method handles arbitrary missing modalities during inference, it still relies on a fixed set of known modalities during training. Adapting to entirely new modalities not seen during pre-training remains an open challenge. 2) Our method pre-trains individual encoders for each modality. While effective, this increases model complexity and limits parameter sharing. Developing a unified encoder architecture that can handle all physiological signals jointly is a promising direction for future work.

\section{Broader Impacts}
\label{broaderimpacts}
\method has the potential to advance multimodal physiological computing and BCI systems by providing a scalable and adaptable foundation model. Its ability to generalize across datasets and tasks while remaining resilient to missing modalities makes it well-suited for real-world scenarios where data completeness and consistency cannot be guaranteed. Applications include assistive healthcare systems, emotion-aware computing, cognitive load monitoring, and sleep analysis. However, care must be taken to address potential privacy concerns when deploying models trained on sensitive physiological data. Future efforts should also explore fairness, energy efficiency, and interpretability to ensure the responsible use of such foundation models in clinical and everyday applications.


\end{document}